\documentclass{statsoc}

\title[Historical battle deaths]{Changepoint analysis of historical battle deaths}
\author{Brennen T. Fagan}
\address{Department of Mathematics, University of York,
York,
UK.}
\email{btf500@york.ac.uk}
\author{Marina I. Knight}
\address{Department of Mathematics, University of York,
York,
UK.}
\author{Niall J. MacKay}
\address{Department of Mathematics, University of York,
York,
UK.}
\author[Fagan {\it et al.}]{A. Jamie Wood}
\address{Departments of Mathematics and Biology, University of York,
York,
UK.}

\usepackage[a4paper]{geometry}
\usepackage{graphicx,natbib,psfrag,epsf,algorithm,algorithmicx}
\usepackage{times}
\usepackage{amsmath,amsfonts,amssymb}
\usepackage{verbatim}
\usepackage{mathrsfs}
\usepackage{enumerate}
\usepackage[xcolor]{changebar} 	    
\usepackage[caption = false]{subfig}
\usepackage{anyfontsize}            
\usepackage{inputenc}

\usepackage{etoolbox}
\makeatletter
\patchcmd{\@makecaption}
  {\parbox}
  {\advance\@tempdima-\fontdimen2} 
  {}{}
\makeatother

\newcommand{\bcb}{\begin{changebar}}
\newcommand{\ecb}{\end{changebar}}

\newif\ifshowbrennen
\showbrennentrue

\definecolor{dgreen}{rgb}{0,0.3,0} 

\begin{document}
\begin{abstract}
It has been claimed and disputed that World War II has been followed by
a `long peace', an unprecedented decline of war.  We conduct a full changepoint
analysis of well-documented, publicly-available battle deaths datasets,
using new techniques that enable the robust detection of changes
in the statistical properties of such heavy-tailed data. 
We first test and calibrate these techniques. 
We then demonstrate the existence of changes, independent of data presentation, at around 1910 and 1950 CE, bracketing the World Wars, and around the 1830s and 1994 CE. 
Our analysis provides a methodology for future investigations and an empirical basis for political and historical discussions.
\end{abstract}

\section{Introduction}\label{sec:intro}
Is war declining?
The record of historical battle deaths surely embodies more human value than any other conceivable dataset, for every unit in every data point is a human life violently taken, yet its structure remains poorly understood.
Pioneering work was done in the \emph{Journals of the Royal Statistical Society} \citep{Richardson44_Distribution,Richardson46_Number,Richardson52_Contiguity,Moyal49_Distribution} by the Quaker pacifist Lewis Fry Richardson.
Richardson discovered one of the few robust quantitative results in political science \citep[p.~143-67]{Richardson60_Statistics}, that deaths in deadly quarrels are well described by two power-law distributions \citep{Clauset09_Power}, with powers of approximately 2.4 from murders up to events with about 1000 dead, and 1.5 for events of more than 1000 dead (`wars') \citep[Figure~4]{Richardson60_Statistics}.
On the question of whether humanity's propensity for deadly violence has fundamentally altered, Richardson's final conclusion was that `the observed variations [in battle deaths] might be merely random, and not evidence of any general trend towards more or fewer fatal quarrels' \citep[p.~141]{Richardson60_Statistics}.
The newly apparent phenomenon of the 60 years since Richardson's book is the post-World War II `long peace', although one might just as well characterize the 20th century by the `great violence' \citep[p.~4]{Clauset18_Trends_and_Fluctuations} of its first half.

Every point of this data takes place in a web of human society, culture and politics.
To analyse this requires a broad sweep of multidisciplinary qualitative analysis, and an astonishing book by Pinker -- suffused with individual statistics, but not overtly a statistical work -- concludes that an individual's likelihood of violent death has greatly declined over the centuries \citep{Pinker11_Better}.
\citet{Goldstein11_Winning} reaches similar conclusions, giving a great deal of recent credit to the United Nations.
The idea of an invariant human tendency towards violence retains its proponents \citep{Huntington89_No,Gray12_Another}, although others who accept the violence of pre-civilized societies (e.g. \citet{Gat13_Declining}) nevertheless stress its amelioration by the continuing development of the Hobbesian state- and super-state-Leviathan.
A classic work by \citet{Gaddis86_Long} lays out multiple possible explanations for the post-World War II absence of large scale war.

The question has become hugely controversial in the last few years, playing out rather publicly in the pages of \emph{Significance}, the joint US/UK magazine for professional statisticians, between Michael Spagat and Stephen Pinker on the one hand and Pasquale Cirillo and Nassim Nicholas Taleb on the other \citep{Spagat15_World,Cirillo16_Significance,Spagat16_Letter}.
\citet{Cirillo16_Statistical} applied techniques from extreme value theory to an unpublished dataset covering 60 CE until 2015 CE and failed to find evidence for any change in arrival time or distribution.
\citet{Clauset18_Trends_and_Fluctuations} arrived at a similar conclusion by applying standard statistical techniques to the publicly available Correlates of War dataset.
Spagat and Pinker found it erroneous to conclude that there was no change in the distribution of violence since World War II without explicit comparison and testing of the periods immediately before and after.
Indeed, they identified several qualitative changes that suggest the world has become more peaceful, in line with \citet{Pinker11_Better}.
In the same vein \citet{Spagat18_Decline} tested the null hypothesis of no change in the magnitude of large wars before and after 1945 or 1950 and found sufficient evidence to reject it for some definitions of large wars.
\citet{Hjort18_Towards} performed a restrictive changepoint analysis limited to a single changepoint and requiring parametric assumptions, and subsequently found 1965 to be the most likely candidate for a change in the sizes of wars.

What has not been done so far, and is the subject of this paper, is a full and comprehensive changepoint analysis of the best-available historical battle deaths datasets.
To conduct a full changepoint analysis on heavy-tailed data, in which extreme data points are `common', is a difficult task for which the methodology has until recently been inadequate.
Our contributions are (i) to calibrate the components of the flexible methodology of \citet{Killick12_Optimal} and \citet{Haynes17_CROPS, Haynes17_Nonparametric} through simulation studies on generated data with traits akin to the historical data, and (ii) to employ the proposed algorithm to infer in a data-driven manner whether there is sufficient historical evidence to support distributional changes.
We do not posit the existence of any fixed changepoint(s).
To do so, after all, might cause us to miss other interesting phenomena in the data, and introduces human bias –- we will not impose a 2019 view of which moments may have been epochal.
In a historical sense, should one or more changepoints be detected, this provides candidates for approximate times at which something changed in the distribution of wars.
If, for example, a changepoint near World War II were detected, following which the distribution yields fewer deadly events, this would lend credence to the `long peace'.

The article is structured as follows.
In Section~\ref{sec:data} we introduce the historical battle deaths datasets.
In Section~\ref{sec:method} we calibrate the relevant methodology, focusing on simulated data and showing that there does indeed exist a changepoint methodology that is successful in identifying statistical changes in power-law distributions.
In Section~\ref{sec:applic} we use this methodology to analyse the historical datasets.
We conclude in Section~\ref{sec:conclusion} with an interpretation and discussion.

\section{Battle deaths datasets}\label{sec:data}

Since the pioneering work of Richardson there have been many attempts to create datasets quantifying violence.
The construction of these datasets raises a number of important questions, first of definition and then also of incomplete or biased knowledge.
\citet[p.~xxxvi,~4--12]{Richardson60_Statistics} was acutely aware of these issues, which is why he chose to focus on `deadly quarrels' of all sizes and types.
More recent approaches to data collection often focus on sub-types of deadly quarrels, such as battle deaths above a set threshold, as in the Correlates of War datasets \citep{Sarkees10_Resort}, or terrorism, as in the Global Terrorism Database \citep[p.~9--10]{START16}.
For recent reviews see \cite{Bernauer12_New} and \cite{Clauset18_Trends_in_Conflict}.

Even if we do settle on an appropriate subset of violence, there are still a number of issues to be decided.
There are complex questions regarding the inclusion of non-combatants, particularly in asymmetric (typically, insurgent) warfare.
An extreme example is the Taiping rebellion in 19th Century CE China.
There is no question that this tragic campaign led to enormous loss of life, but how many of the dead were combatants?
How many civilian deaths have been accounted for? How does one separate battle deaths from those caused by famine and disease and those caused in other, simultaneous rebellions?
Estimates for this particular event vary over at least an order of magnitude.
It is commonly stated that approximately 20 million died in total in the Taiping rebellion \citep{Spence96_Taiping, Reilly04_Taiping, Fenby13_Modern_China}.
\citet{Sivard91_World} indicates 5 million military deaths with 10 million total (in comparison to 300,000 due to simultaneous rebellions) using data due to Eckhardt.
\citet{Worden88_China} reports that 30 million were reported killed over 14 years.
\citet{Platt12_Autumn} reports in the epilogue 70 million dead, along with the standard 20 -- 30 million figure and criticisms of both of these numbers.
\citet{Deng03_Fact} indicates similar numbers from Chinese sources, but notes their interrelation with famine.
However, the Correlates of War dataset reports 26,000 (Chinese), 85,000 (Taipings), 25 (U.K.) battle deaths albeit only for the second, intra state phase of the war.
Battle deaths for the initial, non state phase are listed as unknown.
The Gleditsch dataset is consistent with the Correlates of War values.
Particular difficulty arises where there is disagreement between  contemporary (or even political descendants of) participants, and especially where one or other side has a different level of control or vested interest in the interpretation of the event.

A further issue emerges regarding granularity and data aggregation \citep{Cirillo16_Statistical}.
What constitutes an individual event, and to what extent should individual actions be distinguished within a larger conflict?
For example, should the different fronts in World War II be considered separate?
Should World Wars I and II be considered merely as more active periods within a global conflagration which encompasses both?
This might seem more natural from a Russian or Chinese than from an Anglosphere perspective – for example, how should we handle the Japanese invasion of Manchuria and Sino-Japanese War of 1931-1945, or the Russian civil war of 1917-1922?
And since such events (and related combinations thereof) happen over an extended period, to which point in time should we assign the combined event? Both inappropriate aggregation and inappropriate disaggregation can lead to artefacts \citep{Cristelli12_Power}.
To counter this as much as is possible, we must work only with well-known, publicly available, datasets that handle the data consistently and with clear assumptions on data gathering and aggregation.

\begin{figure}
       \centering
       \includegraphics[width=\linewidth]{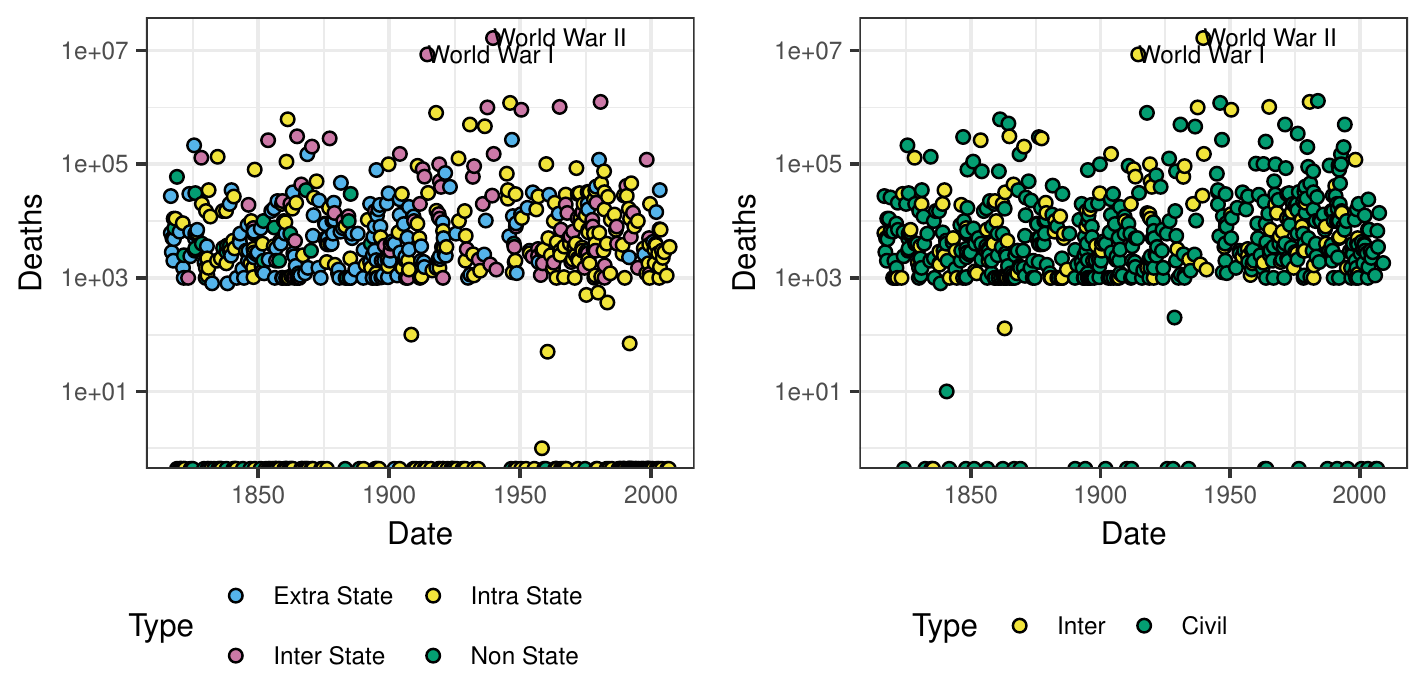}
       \caption{\textbf{Datasets on logarithmic axis.}
       Left: CoW dataset. Right: Gleditsch dataset. The World Wars are labelled for reference.
       Colours indicate the different subsets defined for each dataset and are indicated in the respective legends.
       }
       \label{fig:CoWdata}
\end{figure}

We acknowledge that none of the available datasets is ideal, as each has varying criteria for inclusion of events; and indeed the available historical data themselves are not ideal, due to, for instance, biases in the record.
The two datasets we use are the \emph{Correlates of War} \citep[hereafter,~\emph{CoW}]{Sarkees10_Resort} and a dataset due to \citet[hereafter~the~\emph{Gleditsch~dataset}]{Gleditsch04_Revised}.
We note that the Gleditsch~dataset was originally based upon the CoW~dataset, although divergent evolution has occurred since.
The CoW dataset has four different subsets (inter state, intra state, extra state and non state), whereas the Gleditsch dataset identifies civil and inter state wars.
In our analysis, for simplicity, we consider each event to have occurred at its start date for the purposes of ordering.
We otherwise discard date data, although we are mindful of the possibility of instabilities due to the granularity and uneven temporal distribution.
In Figure~\ref{fig:CoWdata}, we show the CoW dataset, on the left, and the Gleditsch dataset, right, on a logarithmic scale for better visual representation of the data.
For events that are listed but have no value recorded, we present the events on the bottom of the plots at their listed time, but do not include them in the analysis.

A controversial question is whether we should consider the absolute number of deaths caused in a conflict or the number relative to global population, see e.g. \cite{Spagat18_Decline}.
There are good arguments for each choice.
The relative number, favoured by \cite{Pinker11_Better} and \cite{Spagat18_Decline}, approximates the probability of being killed in a particular event, and thus the significance of the event to the average person alive at the time.
On the other hand, we acknowledge the criticisms of \cite{Epstein11_Review} and \citet[p.~16]{Cirillo16_Statistical} that one should not be satisfied merely with a decreasing proportion of battle deaths if the raw values stay high or increase.
We therefore conduct our analyses on both raw data and data normalized by world population, computed using the HYDE 3.2.1 dataset \citep{KleinGoldewijk17_HYDE}.

\section{Methodology for detecting changepoints for power-law distributions}\label{sec:method}

\subsection{Brief review of changepoint detection methodology}
Recall that our aim is to identify in the battle deaths datasets the existence and locations, if any, at which we observe a change in the statistical properties.
For the data of interest in this article, battle deaths, a number of complex issues arise, not only because of the quality of the underlying data but also because the data are characterised by heavy tails and typically modelled using a power-law distribution.

Simply put, a typical changepoint search method consists of three components: an algorithm, a cost function and a penalty.
The combination of cost function and penalty balances explanatory worth against model complexity, valuing the ability of the changepoints to describe the data while penalizing the additional complexity they introduce (usually in proportion to their number).
Often, the cost function is parametric in nature; it assumes some knowledge about how the distribution is parametrised.
This may range from a simple assumption – for example, that the mean or variance exists (e.g. CUSUM, \cite{Page54_Continuous}) – to something more specific, such as that the data follow a normal distribution.

Formally, we denote the time-ordered observations by $y_1, \ldots, y_n$ with potentially $m$ changepoints at (integer) ordered locations $\underline\tau_{1:m} \equiv (\tau_1,\tau_2, \ldots ,\tau_m)$, with $1 \leq m \leq n-1$. Also denote $\tau_0=0$ and $\tau_{m+1}=n$.
The changepoints thus split the data into $m+1$ segments $\{ \underline y_{(\tau_{i-1}+1):\tau_i} \equiv (y_{\tau_{i-1}+1}, \ldots, y_{\tau_i}) \}_{i=1}^{m+1}$ and a cost is associated to each segment, denoted $\mathcal{C}\left( \underline y_{(\tau_{i-1}+1):\tau_i} \right)$ (see, e.g. \cite{Haynes17_CROPS}) .
The penalty function, denoted $f$, aims to control the segmentation size $m$ and contributes to formulating a penalised minimisation problem
\[
\underset{m,\underline \tau_{1:m}}{\mbox{min}}\left(
\sum_{i=1}^{m+1} \mathcal{C}\left(\underline y_{(\tau_{i-1}+1):\tau_i} \right)  + f(m) \right).
\]
Often-encountered cost choices are the negative log-likelihood \citep{Chen00_ParametricCPA} and quadratic loss \citep{Rigaill15_Pruned}.
The penalty is often chosen to be a linear function $f(m)=(m+1)\beta$, with e.g. $\beta=2p$ (Akaike's information criterion or AIC \citep{Akaike74_New}), $\beta=p \log(n)$ (Bayesian information criterion or BIC, also known as Schwarz's information criterion or SIC \citep{Schwarz78_Estimating}), or $\beta=2p \log\log(n)$ \citep{Hannan79_Determination} where $p$ denotes the additional number of parameters introduced by adding a changepoint.

To cope with the heavy tails of battle deaths data we explore the utility of  a nonparametric changepoint analysis.
This was first proposed by \citet{Zou14_Nonparametric} and then incorporated by  \citet{Haynes17_Nonparametric} by means of the empirical distribution (ED) into the dynamic programming algorithm for optimal segmentation search of \citet{Killick12_Optimal} (PELT), thus referred to as ED-PELT.
We explore (ED-)PELT with the classical penalty choices introduced above, but we also consider the modified Bayesian information criterion (mBIC) of \citet{Zhang07_Modified} and the Changepoints for a Range of PenaltieS (CROPS) algorithm of \citet{Haynes17_CROPS} that explores optimal segmentations across a range of penalties in order to bypass the disadvantage of ED-PELT of having to supply a value for $p$.
While ED-PELT \citep{Haynes17_Nonparametric} has been shown to outperform competitor methods when mildly deviating from the usual normal distribution assumption for the observed data, to the best of our knowledge none of the standard methods for changepoint detection (for a recent review see \citet{Truong18_Selective}) has been specifically tested on data obeying power-law distributions. 

\subsection{Simulation study}\label{subsec:sims}
This section performs an in-depth exploration of the performance of existing segmentation methods for simulated data following power-law distributions with powers akin to those documented for historical battle deaths. The wide pool of candidate methods is first narrowed down in Section~\ref{subsubsec:initial}, and the thorough testing in the subsequent sections leads us to propose a changepoint detection algorithm (Algorithm~\ref{alg:cptalgo} in Section~\ref{subsubsec:final}) suitable for our context.

In order to compare methods, we consider three metrics: the Hausdorff metric, the adjusted Rand index (henceforth, ARI), and the true detection rate (henceforth, TDR).
The first measures segmentation by reporting the worst minimum distance between two points in the true and discovered changepoint sets \citep{Truong18_Selective}.
The Rand index measures (cluster) accuracy by comparing the relationships of data in each cluster in the discovered changepoint set to the true, \citep{Truong18_Selective}.
We use the adjusted Rand index, implemented in \texttt{mclust}, to account for clustering due to chance \citep{Scrucca17_mclust}.
Total agreement between clusters results in an ARI of 1, while the expected value of a random partition of the set is 0.
Finally, the true detection rate gives us an understanding of how many changepoints detected are true or false by checking to see if a true changepoint happened near a detected one \citep{Haynes17_Nonparametric}.
A TDR of 1 indicates that every changepoint detected is within a given distance of at least one true changepoint, while a TDR of 0 indicates that every changepoint is outside such a distance.
First, for direct comparison, we consider a radius of acceptance of 0 \citep{Haynes17_Nonparametric}.
In order to choose appropriate further radii, we consider the historical data.
For example, World War I could have conceivably occurred two years earlier due to conflicts in the Balkans.
On one side of $67.0\%$, $78.4\%$, and $77.9\%$ of wars, 3, 5, and 8 new wars will have occurred within 1, 2, or 3 years respectively.
Hence, we use radii of 3, 5, and 8 to roughly represent 1, 2, or 3 years in the historical data set.
We also do not include the endpoints of the data as changepoints for this calculation.

All simulation tests were carried out in R. In particular, data generation was performed using the \texttt{poweRlaw} R-package \citep{Gillespie17_poweRlaw}, while changepoint analyses were carried out using the \texttt{changepoint} \citep{Killick16_changepoint} and \texttt{changepoint.np} \citep{Haynes16_changepoint.np} R-packages.
As the name suggests, the extension $\star$\texttt{.np} in the package name and associated function stands for the nonparametric approach of \citet{Haynes17_Nonparametric}. Visuals were compiled using the \texttt{ggplot2} R-package \citep{Wickham16_ggplot2}.

\subsubsection{Initial method screening}\label{subsubsec:initial}

To benchmark the various candidate methods, we first screened the possible combinations of cost and penalty corresponding to different data modelling distributions.
Table \ref{tab:cpt}  summarises the available functions and options, as implemented in the changepoint packages above, while noting restrictions on combinations of methods.
Some of the arguments provided require additional information which we set to be the same across all tests.
Specifically: the type I error probability was set to 0.05; the penalty range for CROPS was set to $10^0$--$10^6$; the maximum number of segments in SegNeigh \citep{Auger89_Algorithms} was set to 61, and the maximum number of changepoints required by BinSeg \citep{Scott74_Cluster} was set to 60.

\begin{table}
	\caption{\label{tab:cpt}Function options for \texttt{changepoint} and \texttt{changepoint.np} R-packages.
	The first column corresponds to the R function used, while the other three correspond to arguments that determine how the analysis is performed.
	Note that not every combination of options within a function are valid: SegNeigh \citep{Auger89_Algorithms} cannot be used with mBIC; PELT, mBIC and Asymptotic cannot be used with CUSUM; PELT and mBIC cannot be used with CSS \citep{Inclan94_Use}; Asymptotic cannot be used with Poisson; CROPS was designed for use in conjunction with PELT.
	In particular, \texttt{cpt.np} is particularly restricted.
	}
	
	\centering
	\begin{tabular}{|l|l|l|l|} \hline
		Function 			& \texttt{penalty} 	& \texttt{method} 	& \texttt{test.stat} \\ \hline
		\texttt{cpt.mean}	& SIC/BIC			& AMOC              & Normal			 \\
		\texttt{cpt.var}	& mBIC				& PELT 				& CUSUM	(\texttt{cpt.mean} only)\\
		\texttt{cpt.meanvar}& AIC 				& SegNeigh          & CSS (\texttt{cpt.var} only)	\\
		            		& Hannan-Quinn 		& BinSeg            & Exponential (\texttt{cpt.meanvar} only)\\
							& Asymptotic 		&					& Poisson (\texttt{cpt.meanvar} only)\\
							& CROPS 			&					&				 \\ \hline
		\texttt{cpt.np}  	& SIC/BIC			& PELT              & Empirical Distribution \\
		                    & mBIC				&  				    & \\
		                    & AIC 				&                   & 	\\
				            & Hannan-Quinn 		&                   & \\
							& CROPS		        &					& \\ \hline
	\end{tabular}
\end{table}

We assessed segmentation outcomes across $N = 1000$ trials with data of length $n = 600$ featuring a single changepoint ($m = 1$) located at $\tau_1 = 300$.
The first segment consisted of data simulated from a power-law distribution with parameter $\alpha=2.05$, while for the second segment we chose $\alpha=2.55$.
Across our simulations we set the lower cutoff for the power-law to hold, to be 10. 

\begin{figure}
	\centering
	\includegraphics[width=0.8\linewidth]{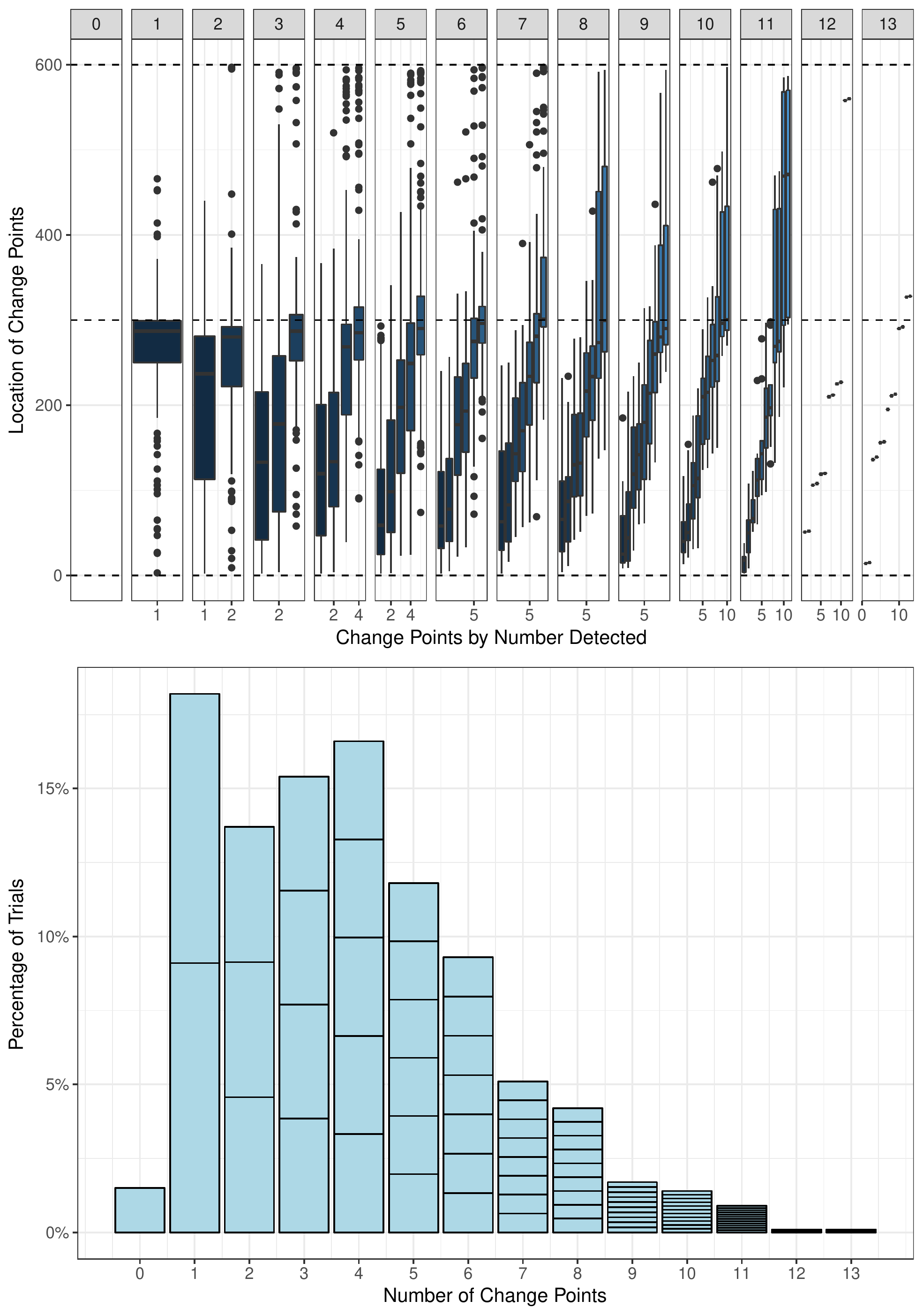}
	\caption{\textbf{Test case with better than average behaviour.} Segmentation generated using \texttt{cpt.meanvar} with BinSeg, mBIC and an exponential distribution. Whilst there are good aspects to this finding, the method commonly overfits and tends to assume changepoints happen in the $\alpha = 2.05$ segment. This combination has median Hausdorff of 190, median ARI of 0.65, and TDR 0: 0.01, 3: 0.05, 5: 0.07, 8: 0.09.}
	\label{fig:1st_Test_Okay}
\end{figure}

\begin{figure}
	\centering
	\includegraphics[width=0.8\linewidth]{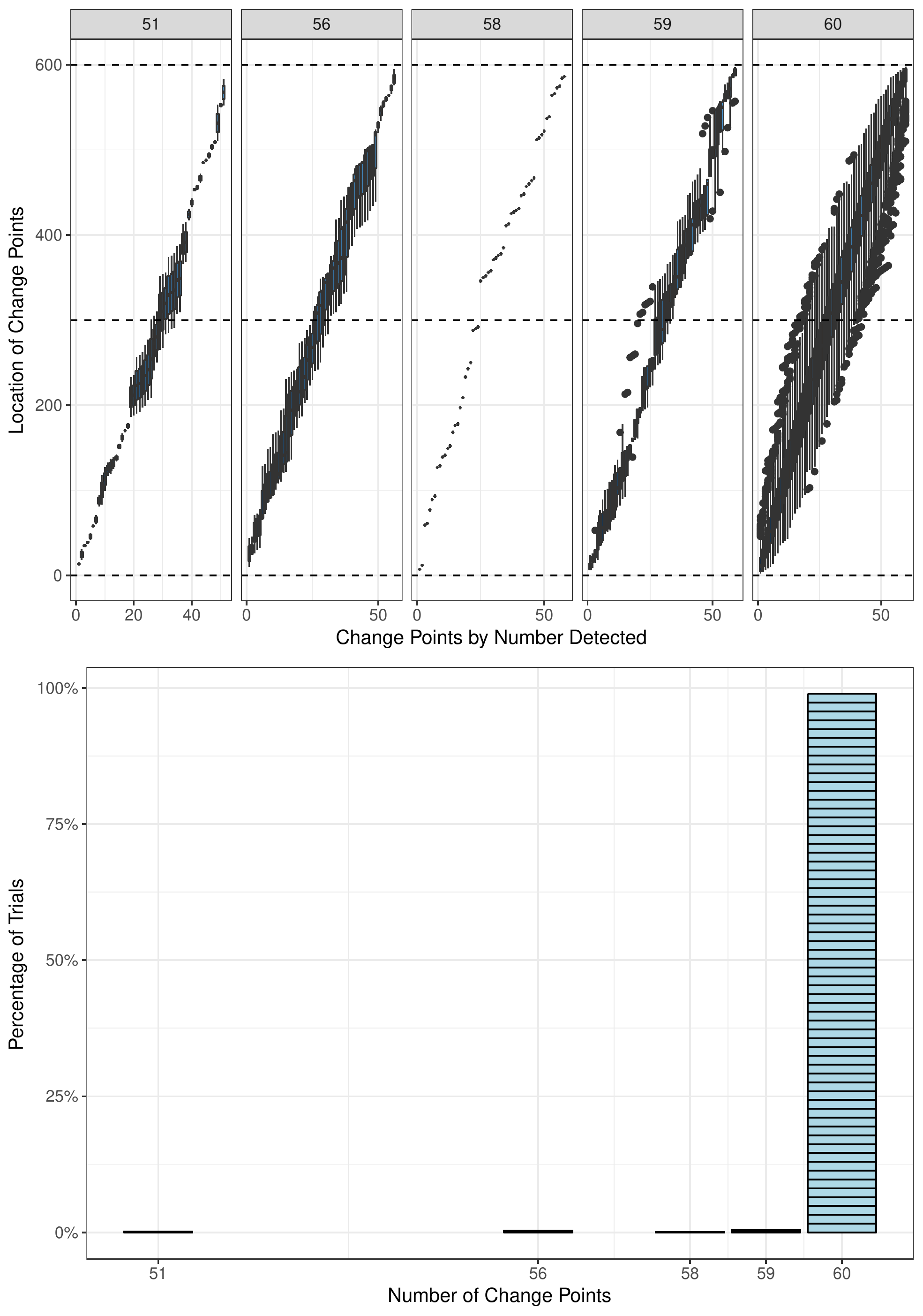}
	\caption{\textbf{Test case with worst behaviour.} Segmentation generated using \texttt{cpt.meanvar} with SegNeigh, an asymptotic penalty and a normal distribution. Results such as this occur with many combinations, and can be regarded as failures. Many combinations result in more than 10 false positives and are only stopped by the maximums provided. This combination has median Hausdorff of 294, median ARI of 0.07, and TDR 0: 0.00, 3: 0.01, 5: 0.01, 8: 0.01.}
	\label{fig:1st_Test_Fail}
\end{figure}

\begin{figure}
	\centering
	\includegraphics[width=0.8\linewidth]{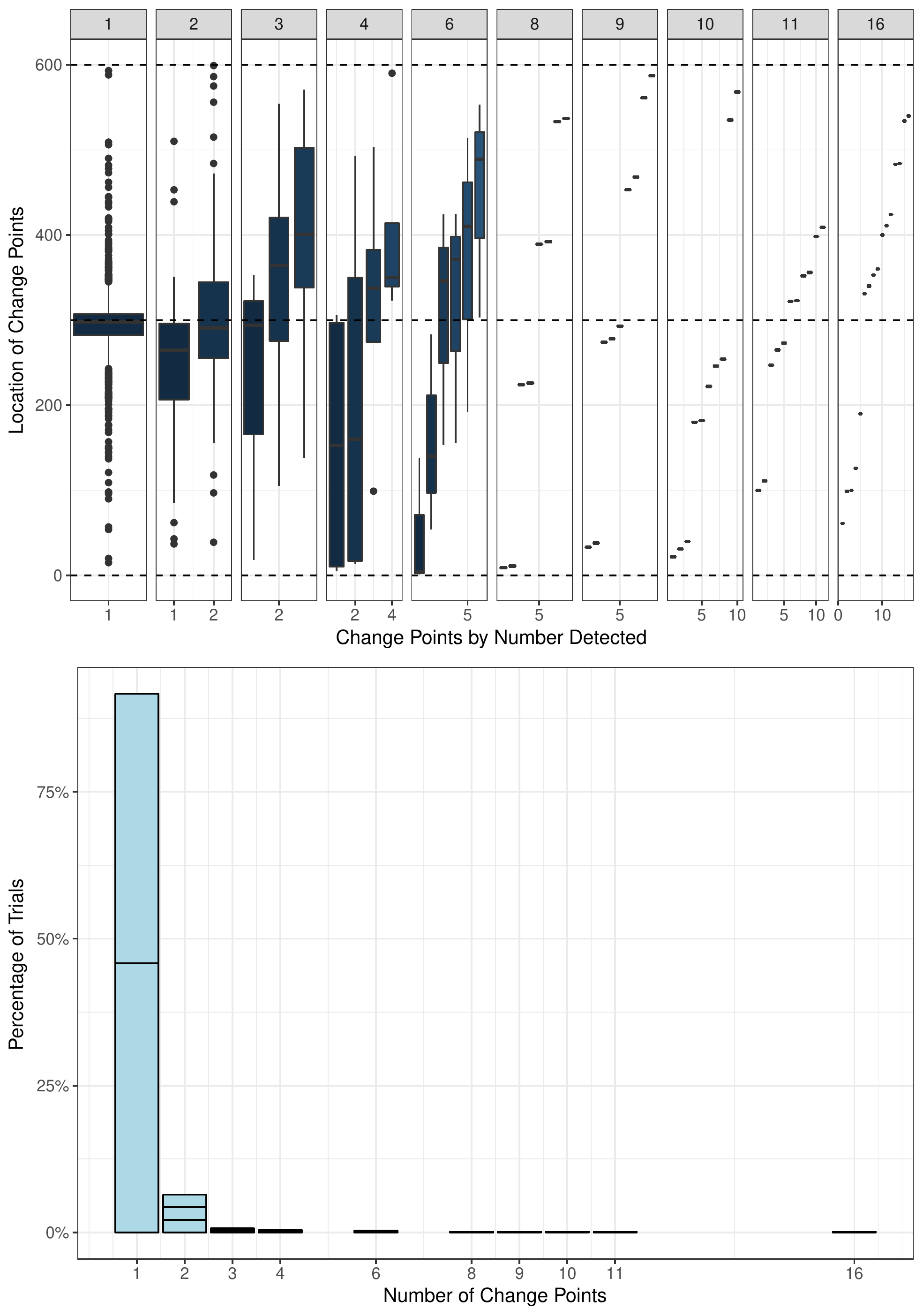}
	\caption{\textbf{Test case with best behaviour.} Segmentation generated using \texttt{cpt.np} with ED-PELT and CROPS and shows some of the best achievable behaviour. Although qualitatively similar to the top sub-plot of Figure \ref{fig:1st_Test_Okay}, there is improved accuracy in the positioning of the changepoints and improved precision and accuracy in the number of points so detected. This combination has the lowest median Hausdorff of 15, highest median ARI of 0.91, and highest TDR 0: 0.04, 3: 0.18, 5: 0.25, 8: 0.32.}
	\label{fig:1st_Test_Success}
\end{figure}

\begin{figure}
	\centering
	\includegraphics[width=0.8\linewidth]{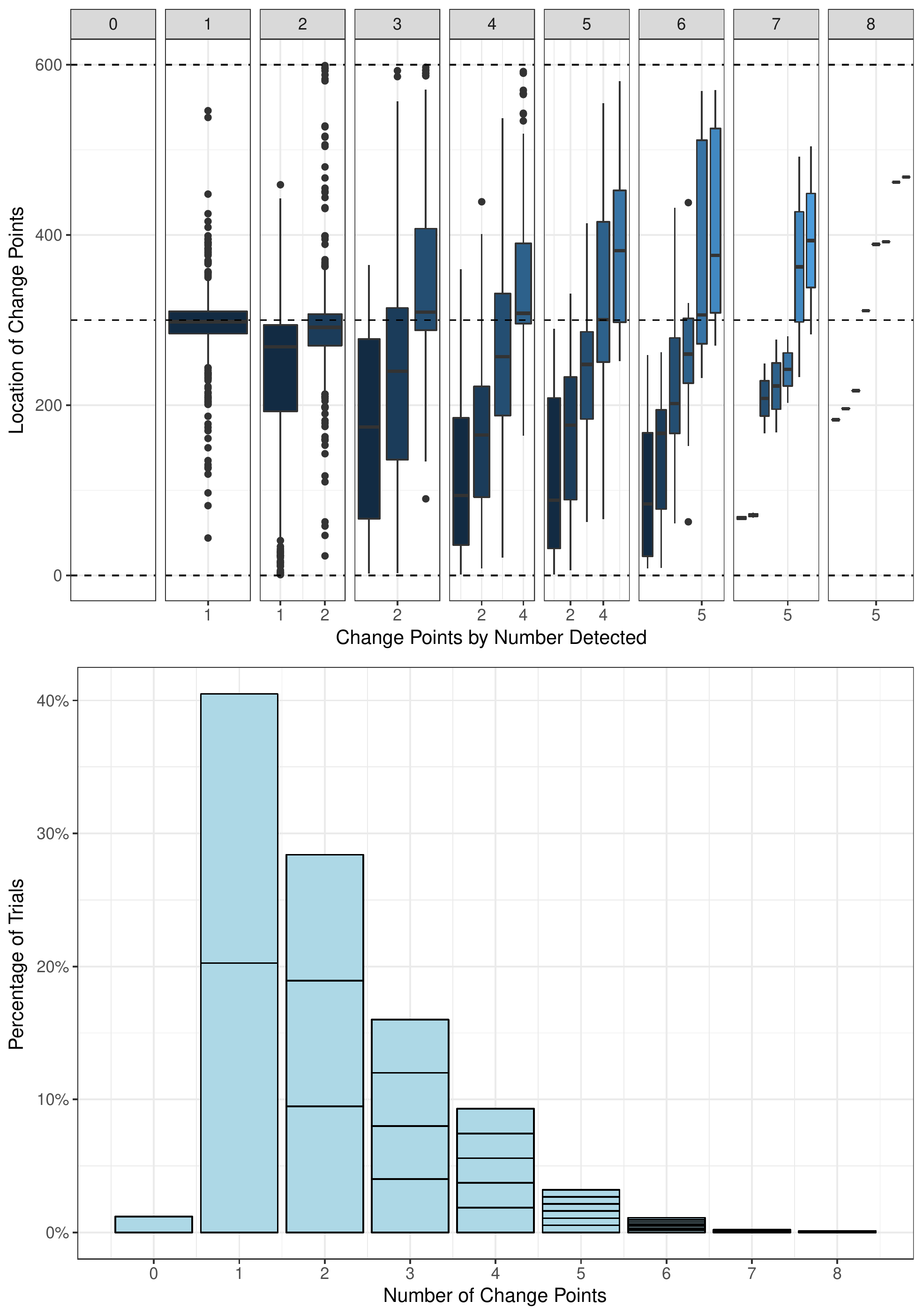}
	\caption{\textbf{Test case with second best behaviour.} Segmentation generated using \texttt{cpt.np} with ED-PELT and mBIC. While not as good at detecting changepoints as CROPS, \texttt{cpt.np} with ED-PELT and mBIC still shows strong potential. This combination has median Hausdorff of 44, second highest median ARI of 0.81, and TDR 0: 0.02, 3: 0.10, 5: 0.14, 8: 0.18.}
	\label{fig:1st_Test_MBIC}
\end{figure}

Figures \ref{fig:1st_Test_Okay} -- \ref{fig:1st_Test_MBIC} give illustrative examples of the types of behaviour of the analyses conducted.
The bottom subplot of each plot indicates the proportion of trials in which a given number of changepoints was detected by the analysis.
The top subplots are arranged by the number of changepoints found and use boxplots to show the location of each changepoint so found.
The middle dashed line is placed along the changepoint.
Across the tested combinations, most failed to identify that there was only a single changepoint, let alone to pinpoint its precise location.

We also note that Figures \ref{fig:1st_Test_Okay} -- \ref{fig:1st_Test_MBIC} do not showcase all possible outcomes.
For example, some combinations result in approximately correct numbers of changepoints but incorrect locations. Even when using \texttt{cpt.np} overfitting is still common with penalties such as AIC or BIC.
PELT and CROPS are also no guarantee of success; \texttt{cpt.mean} with PELT, CROPS, and a normal distribution results in preferential selection for even numbers of changepoints, overfitting, and placement in the middle of the $\alpha = 2.05$ segment.
Of the \texttt{changepoint} methods, `at most one changepoint' \citep[henceforth, AMOC]{Page54_Continuous} was most successful, as it was tied with itself for second lowest median Hausdorff measure (39), third highest median ARI (0.76), and second highest TDR (0: 0.03, 3: 0.12, 5: 0.16, 8: 0.20).
Due to its obvious restriction, we felt compelled to discard AMOC however.
Based on these findings, we therefore select ED-PELT with CROPS and mBIC to continue with (implemented under function \texttt{cpt.np} in the \texttt{changepoint.np} package). We find appealing not only their comparatively strong behaviour but also the theoretical lack of parametric assumptions, suitable for our context.

\subsubsection{Investigation in the presence of at most one changepoint}\label{subsubsec:mid}

In order for our explorations to be relevant to the real battle deaths data, we choose power-law exponents ($\alpha$) close in value to Richardson's law, and test the segmentation robustness against numerical proximity, order and false positive detection, as detailed in Table \ref{tab:General_Tests_1}.
In general, we found that ED-PELT performs well with both CROPS and mBIC penalties, but with CROPS outperforming mBIC in most cases.
Both benefit from increased segment lengths with increased precision of number of changepoints detected and increased ARI (in contrast to the other penalty options, which claim more changepoints occur as segment lengths increase).
Performance for both is consistent regardless of exponent and order.
However, mBIC does outperform CROPS in one notable situation: when the two distributions are very close, such as $\alpha_\text{mod} \leq 0.05$, or coincide (no changepoint).
When this occurs, CROPS has a tendency to dramatically overfit the number of changepoints whereas mBIC is more likely to correctly report no changepoints.

\begin{table}
	\caption{\label{tab:General_Tests_1}\textbf{Test Set Parameters.} Each column represents a parameter and its options for the simulated data in Section~\ref{subsubsec:mid}. Each test was performed with $N = 1000$ trials with $m = 1$ changepoint(s) located at $\tau_1 = s$, where $n = 2s$ with generated data with power-law parameters $(\alpha \pm \alpha_{\text{mod}})$ and $(\alpha \mp \alpha_{\text{mod}})$ on each segment.}
	\centering
	\begin{tabular}{|l|l|l|l|} \hline
		Exponent ($\alpha$) 	& Exponent Modifier ($\alpha_\text{mod}$) & Order 		& Segment Length ($s$) \\ \hline
		1.7				& $0$				& Low -- High	& 30			\\
		2.3				& $\pm 0.05$		& High -- Low	& 100			\\
						& $\pm 0.15$		&				& 300			\\
						& $\pm 0.25$		&				& 1000			\\
						& $\pm 0.5$			&				&				\\ \hline
	\end{tabular}
\end{table}

\subsubsection{Investigation in the presence of several changepoints}\label{subsubsec:final}

\begin{figure}
       \centering
       \includegraphics[width=0.68\linewidth]{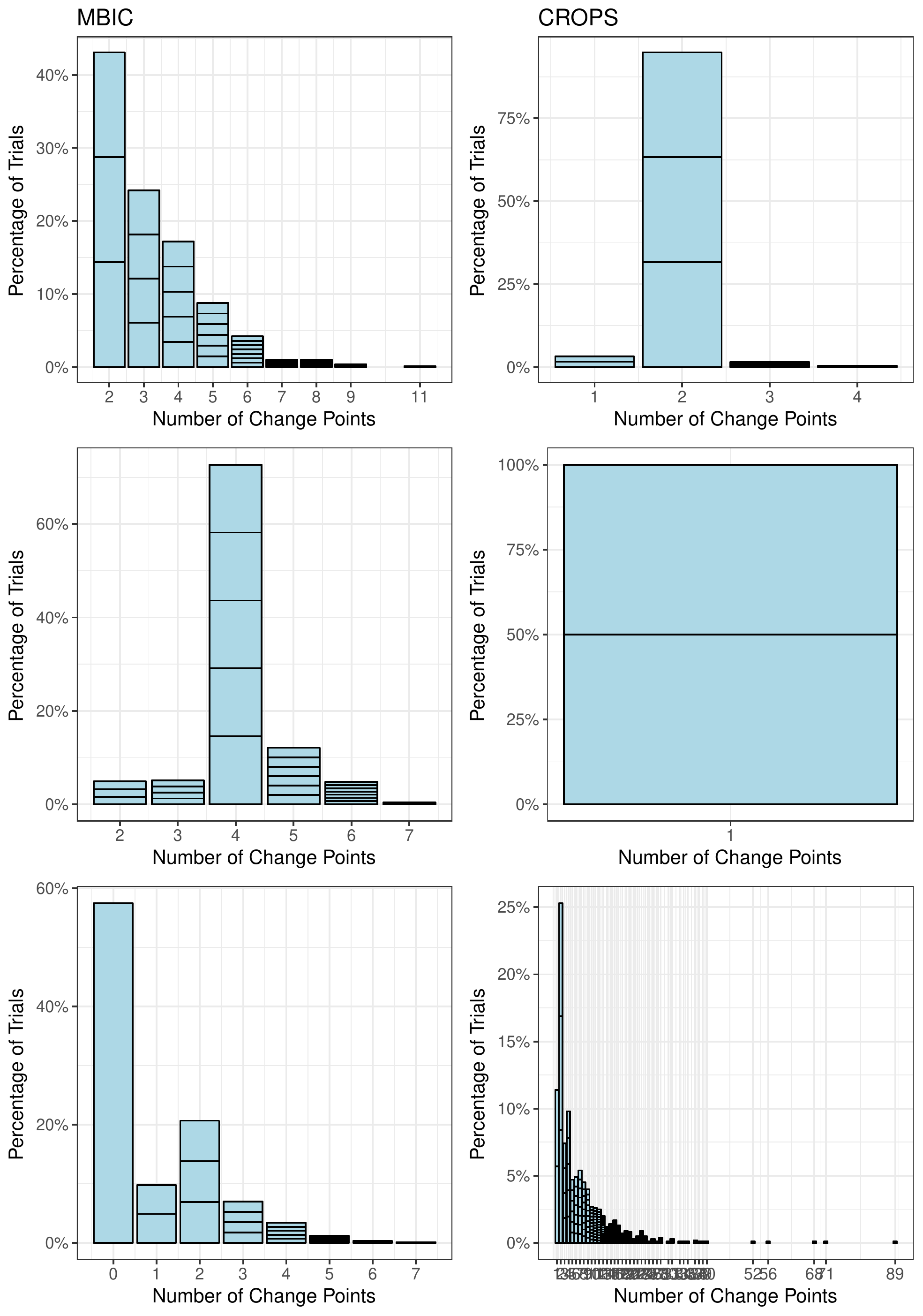}
       \caption{\textbf{Examples comparing behaviour of CROPS and mBIC.} Each row is a different scenario using power-law distributions.
       The sequence length is $n = 1000, 575$ and $600$ respectively.
       In the first row, two changepoints, marking the power-law exponent change from 2.3 to 1.7 to 2.1, are present, and CROPS gives a more accurate result. Simulations show this is the common pattern.
       In the second case four changepoints are present, transitioning across exponents 2.87, 1.83, 2.49, 1.67, and 1.06. CROPS detects only a single changepoint with high precision.
       mBIC outperforms CROPS in this uncommon case.
       Last, we provide a case with no changepoints with an exponent of 1.7, for which CROPS has pathological behaviour, while mBIC succeeds with reasonable precision and accuracy. Note that the behaviour of CROPS is due to a known feature: \citet{Haynes17_Nonparametric} recommend choosing the optimal number of changepoints for CROPS such that it maximises the estimated curvature of the penalty as a function of the number of changepoints. This naturally truncates the data over which the curvature is estimated, removing the possibility of obtaining 0 changepoints on a potentially flat line.
}
       \label{fig:simresults}
\end{figure}

We now expand our investigations beyond the presence of at most one changepoint and explore the outcomes obtained when the data feature several (specifically, two, four or eight) changepoints controlled for variable segment length and data granularity.
Figure \ref{fig:simresults} shows some representative results of each procedure.

In general, our previous findings extend to the multiple changepoints case, as can be seen in the first row of Figure~\ref{fig:simresults}, where CROPS proves to be more precise and accurate in its identification of changepoints (a high TDR), although sometimes too conservative.
The second row of Figure~\ref{fig:simresults} illustrates an uncommon case in which the change across one particular changepoint is so drastic that CROPS identifies it as the only change, missing the less pronounced changes.
In contrast, mBIC mostly successfully identifies these changepoints, showcased in higher ARI's and lower Hausdorff distances.
This uncommon case is more likely to occur when there are a large number of true changepoints (e.g. 8) and small segment sizes.
Finally, the third row of Figure~\ref{fig:simresults} shows the already mentioned case where there is no changepoint in the data; mBIC can detect this reasonably well, but CROPS dramatically overfits.
Subsequent results should be viewed through the lens of these limitations.

We thus conclude that one cannot rely solely on one penalty, but must use the joint findings of CROPS and mBIC to assess the presence of changepoints.
The combined use of the two methods gives good confidence in accurately detecting the correct number of changepoints, as well as their location.
CROPS findings are accurate when small numbers of changepoints are found, whereas changepoints found only by mBIC should be viewed with caution.
Of particular note is that mBIC appears to have an extremely low false negative rate: if mBIC does not find a particular break in the data, then we may be confident that no changepoint is present.
Where mBIC and CROPS agree on identified changepoints, we have a high degree of confidence that this marks a real change of distribution in the data.

In the light of the results above, we propose the following changepoint detection algorithm (Algorithm~\ref{alg:cptalgo}) to employ on the real battle deaths datasets.
This protects against the pathological CROPS case, resulting in increased TDR when considering the changepoint intersection set, while also allowing for a more liberal interpretation of the union of detected changepoints.

\begin{algorithm}
\caption{Proposed changepoint detection algorithm for power-law distributions. (Note $| \cdot |$ denotes cardinality.)}
\label{alg:cptalgo}
\begin{algorithmic}
	\item{Given the time-ordered observations $\underline{y} = \left\lbrace y_1, \ldots ,y_n \right\rbrace$,} segment $\underline{y}$ by applying ED-PELT with penalty
	\begin{enumerate}
		\item{\em mBIC}; denote the estimated set of changepoints as $\underline{\tau}_{\text{mBIC}}$;
		\item{\em CROPS}; denote the estimated set of changepoints as $\underline{\tau}_{\text{CROPS}}$.
	\end{enumerate}
	\item If $|\underline{\tau}_{\text{mBIC}}| = 0$ and $|\underline{\tau}_{\text{CROPS}}| > 2$, then $m = 0$ and the changepoint set is $\underline{\tau} = \emptyset$.
	\item Else
	\begin{enumerate}
		\item Set $\underline{\tau} = \underline{\tau}_{\text{mBIC}} \cap \underline{\tau}_{\text{CROPS}}$ and $m = |\underline{\tau}|$.
		\item For $(\underline{\tau}_{\text{mBIC}} \cup \underline{\tau}_{\text{CROPS}}) \setminus \underline{\tau}$, interpretation is required.
	\end{enumerate}
\end{algorithmic}
\end{algorithm}

\section{Changepoint analysis of historical battle deaths}\label{sec:applic}

\begin{figure}
       \centering
       \includegraphics[width=\linewidth]{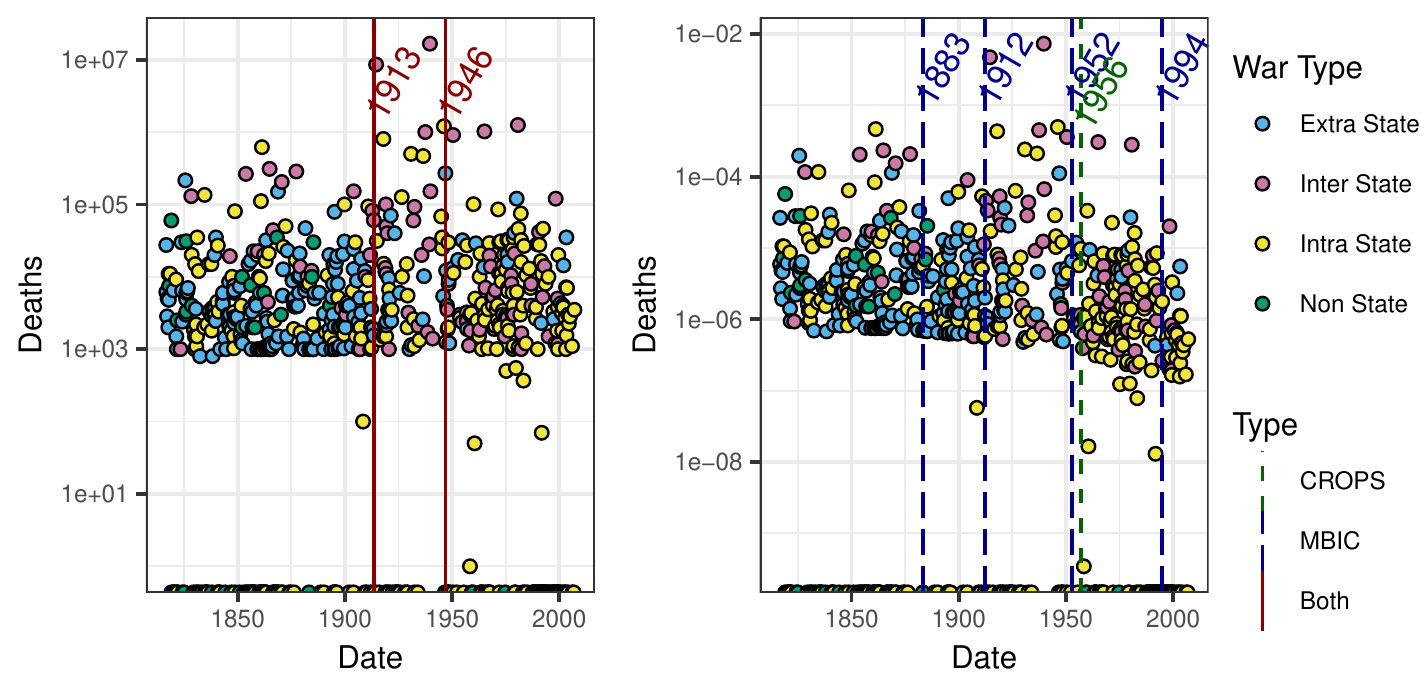}
       \caption{\textbf{Results from applying Algorithm~\ref{alg:cptalgo} to CoW for all data subsets.} On the left we use raw data; on the right, data rescaled by world population at the time of the conflict. Vertical bars indicate detected changepoints annotated by exact years for clarity.}
       \label{fig:cowresults}
\end{figure}

Using the insights gained through the simulation study above, we now apply the proposed algorithm to the publicly-available datasets described in Section~\ref{sec:data}. 
The results indicate with confidence the existence of changepoints in the data. In the raw CoW dataset, shown in Figure~\ref{fig:cowresults}, there are two changes, just prior to World War I and just after World War II.
When scaled by population two more candidate changepoints emerge, in the late 19th century (1883) and in 1994 (and the post-World War II point shifts slightly), but there is less confidence in the changepoints overall since the results are not identical across CROPS and mBIC.
This supports the proponents of the long peace hypothesis, albeit via an argument for what Clauset termed the `great violence' \citeyearpar[p.~4]{Clauset18_Trends_and_Fluctuations}.

\begin{figure}
       \centering
       \includegraphics[width=\linewidth]{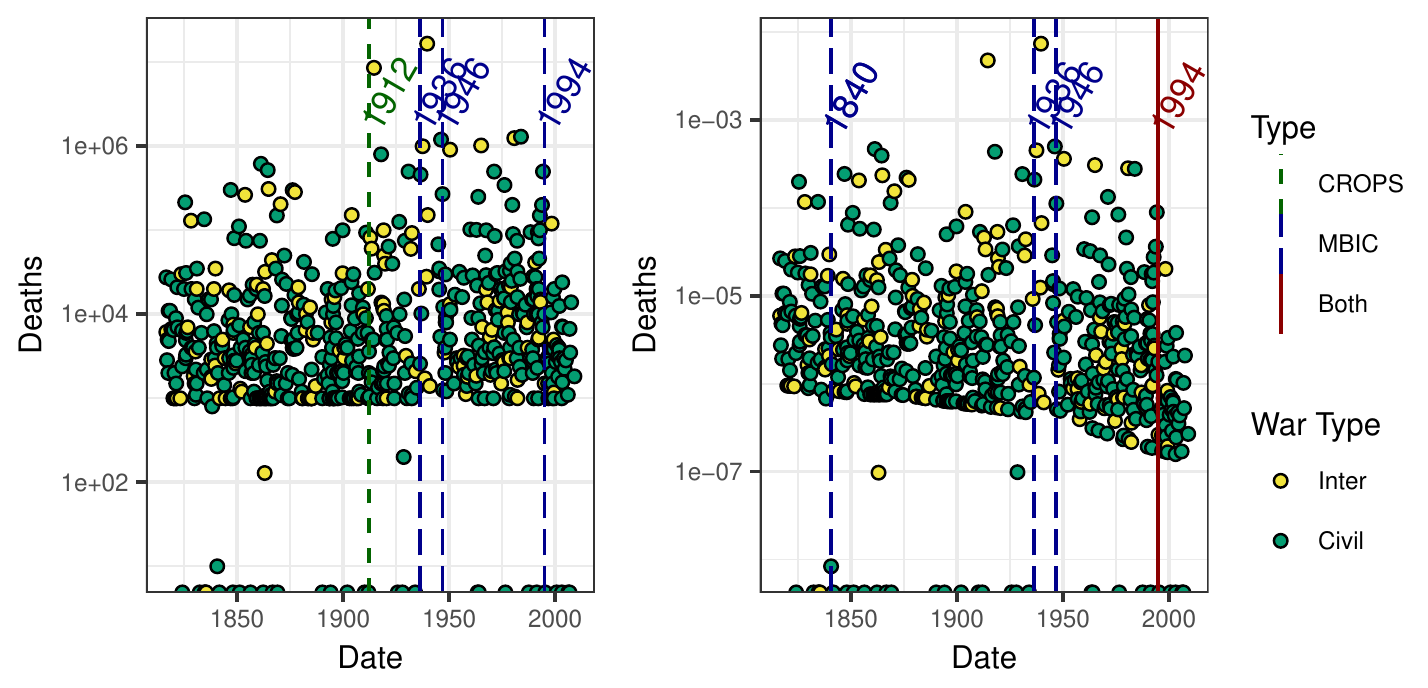}
       \caption{\textbf{Results from applying Algorithm~\ref{alg:cptalgo} to Gleditsch's combined datasets.} On the left we use raw data; on the right, data rescaled by world population at the time of the conflict. Vertical bars indicate detected changepoints annotated by exact years for clarity.}
       \label{fig:gledresults}
\end{figure}

It is less clear to assign changepoints in the Gleditsch raw dataset, but the emerging 1994 changepoint in data scaled by population size is now conclusively found (Figure~\ref{fig:gledresults}). The broad message is similar, with candidate changepoints found pre-World War I, post-World War II, and 1994.
In contrast to the raw CoW dataset, we find evidence for an early 1840 changepoint, similar to the CoW normalised dataset.
In addition, the Gleditsch analysis suggests a change in the mid-1930s, presumably due to different classification of data emerging from the complex civil and proxy wars that took place around this time.

\begin{figure}
       \centering
       \includegraphics[width=\linewidth]{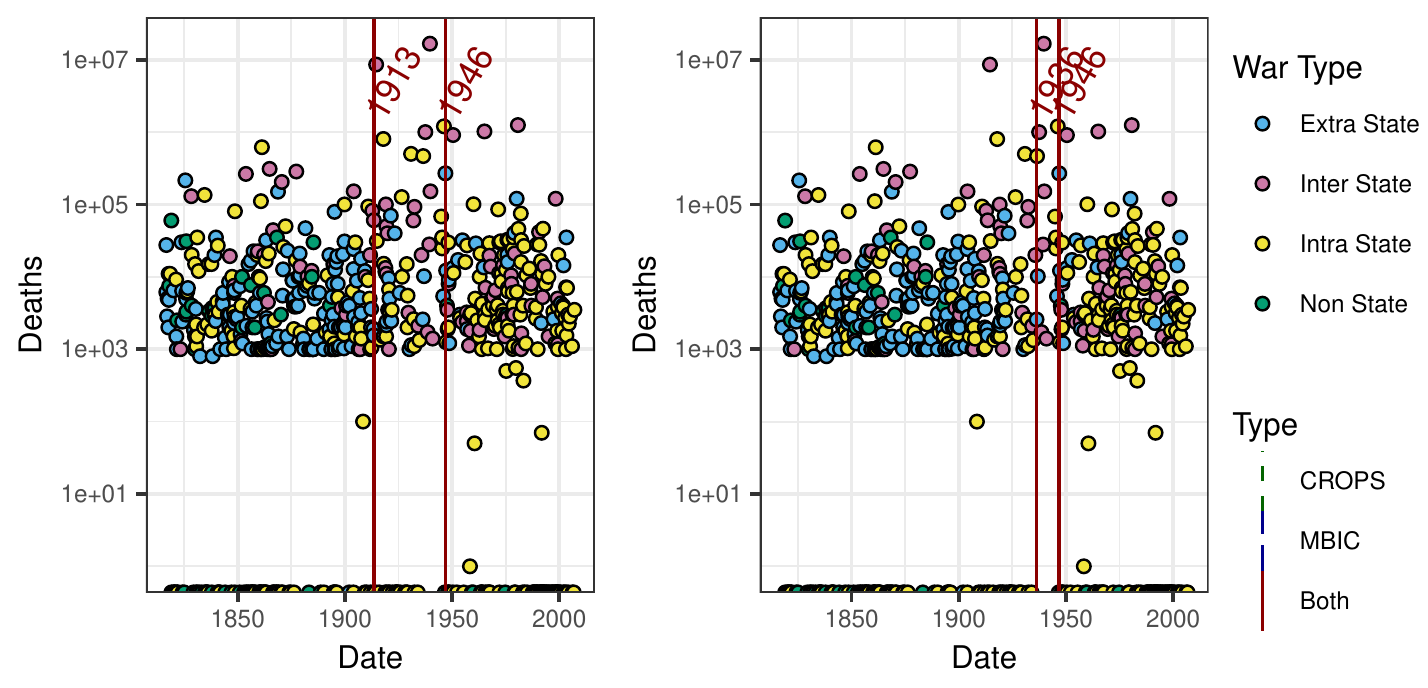}
       \caption{\textbf{Results from applying Algorithm~\ref{alg:cptalgo} to the combined CoW dataset, rescaled as recommended by \citet[p.~30,~32]{Cirillo16_Statistical}.} On the left, the data is rescaled using the current (2018) world population. On the right, data is rescaled using the world population at the time of the conflict. Vertical bars indicate detected changepoints annotated by exact years for clarity.}
       \label{fig:cowrescaled}
\end{figure}

The suggestions made by \citet[p.~30,~32]{Cirillo16_Statistical} to transform the data in order to account for the finite upper bound appear to have little impact (see Figure~\ref{fig:cowrescaled}).
Neither transforming the data to impose a size limit of the 2018 world population on any single war, nor doing so with each event bounded by population at the time of the war, typically changes the number or location of changepoints, especially in the Gleditsch dataset.
Among CoW and its various subsets, an exception is the combined CoW dataset, as shown in Figure~\ref{fig:cowrescaled}.
The limited sensitivity to such transformations is probably due to the lack of data points located sufficiently far in the tail of the distribution –- no single war results in the death of a high proportion of world population.
These results do suggest some sensitivity within the CoW combined dataset, in that the 1913 changepoint in the raw data has a similar likelihood of being identified as the 1936 changepoint in the transformed data.

\begin{figure}
       \centering
       \includegraphics[width=0.6\linewidth]{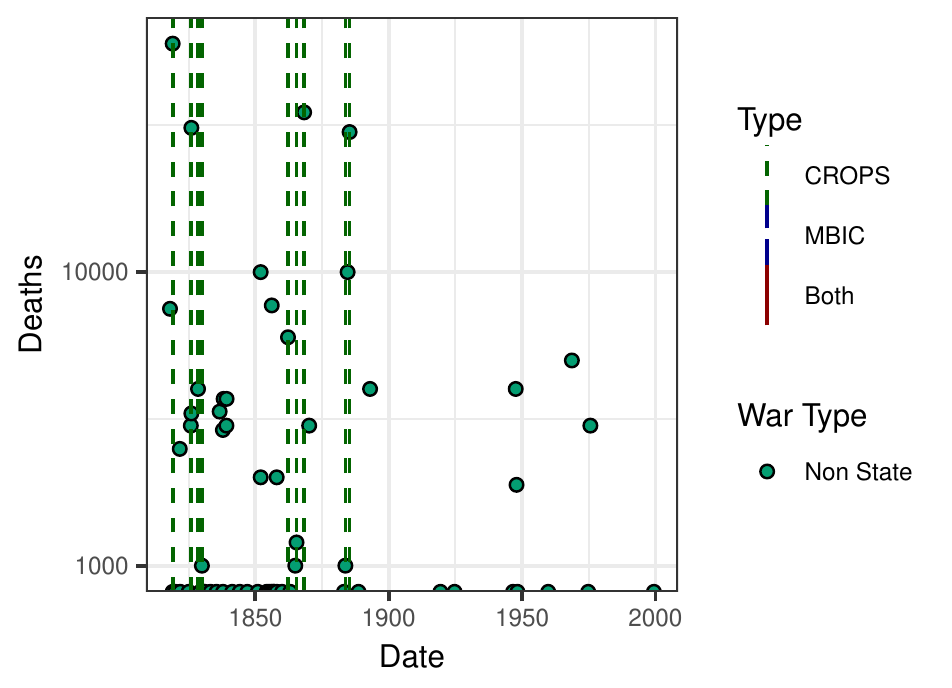}
       \caption{\textbf{Results from applying Algorithm~\ref{alg:cptalgo} to the non state CoW dataset.} No changepoints are found by the mBIC penalty and CROPS finds apparently a large number of tightly cluster points. Note that this result is extremely indicative of no changepoints. For comparison, see Figure~\ref{fig:simresults}.}
       \label{fig:cownonstate}
\end{figure}

An important consistency check is whether any real datasets exhibit no changepoints.
We recall that we have already demonstrated that the no-changepoints case for our methodology is evidenced by a particular combination of a large number of (false) positives from CROPS and few or no (false) positives from mBIC for a wide range of data points, seen in Figure~\ref{fig:simresults}.
Whilst we have established the robustness of the methods against artificial data, the existence of a dataset with no changepoints would clearly help validate our methods while also identifying a setting consistent with the null hypothesis of no change in the statistical properties.
It is therefore worthy of comment that such a dataset within CoW does exist: the CoW non state dataset, shown in Figure~\ref{fig:cownonstate}, has a response clearly of the same type as in the bottom row of Figure~\ref{fig:simresults}, indicating a potential unchanging underlying mechanistic reason for this phenomenon.

To get a sense of the robustness of our approach and to represent the overall prevalence of changepoints, in Figure~\ref{fig:meta} we present an internal meta-analysis across all the analyses we have performed on the CoW and Gleditsch datasets.
This figure shows where changepoints are found in all composing internal data subsets by the proposed algorithm identified in Section~\ref{subsec:sims} (see Figure~\ref{fig:simresults}). In the top panel of each sub-figure, we place a kernel density estimate of the locations of changepoints; the sub-figures and density estimates were created using a 1/5th adjustment to the default bandwidth to sharpen the location of changepoints. In the bottom panel of each sub-figure, we present the data subsets as a time line. Shaded regions, for changepoints over a period of time, and dotted lines, for changepoints located at a single time, indicate the location of clusters of changepoints clustered using the k-means algorithm \citep{Wang2011_Ckmeans}. The area under the density estimation curve is therefore a rough aggregate measure of the likelihood of a changepoint during the period, independent of the magnitude of the change.  This panel gives a clear sense of the robustness (notably 1994), approximate robustness (point around 1830) and the variations that exist in the period 1910-1945 of the changepoints. The graph shows the location of individual points but also more finely-grained variation where multiple methods and datasets produce changepoints at approximately the same point in time. The R-package \texttt{Ckmeans.1d.dp} by \cite{Wang2011_Ckmeans} was used for clustering in this context.

\begin{figure}
       \centering
       \subfloat[CoW raw data]{
           \includegraphics[width=0.85\linewidth]{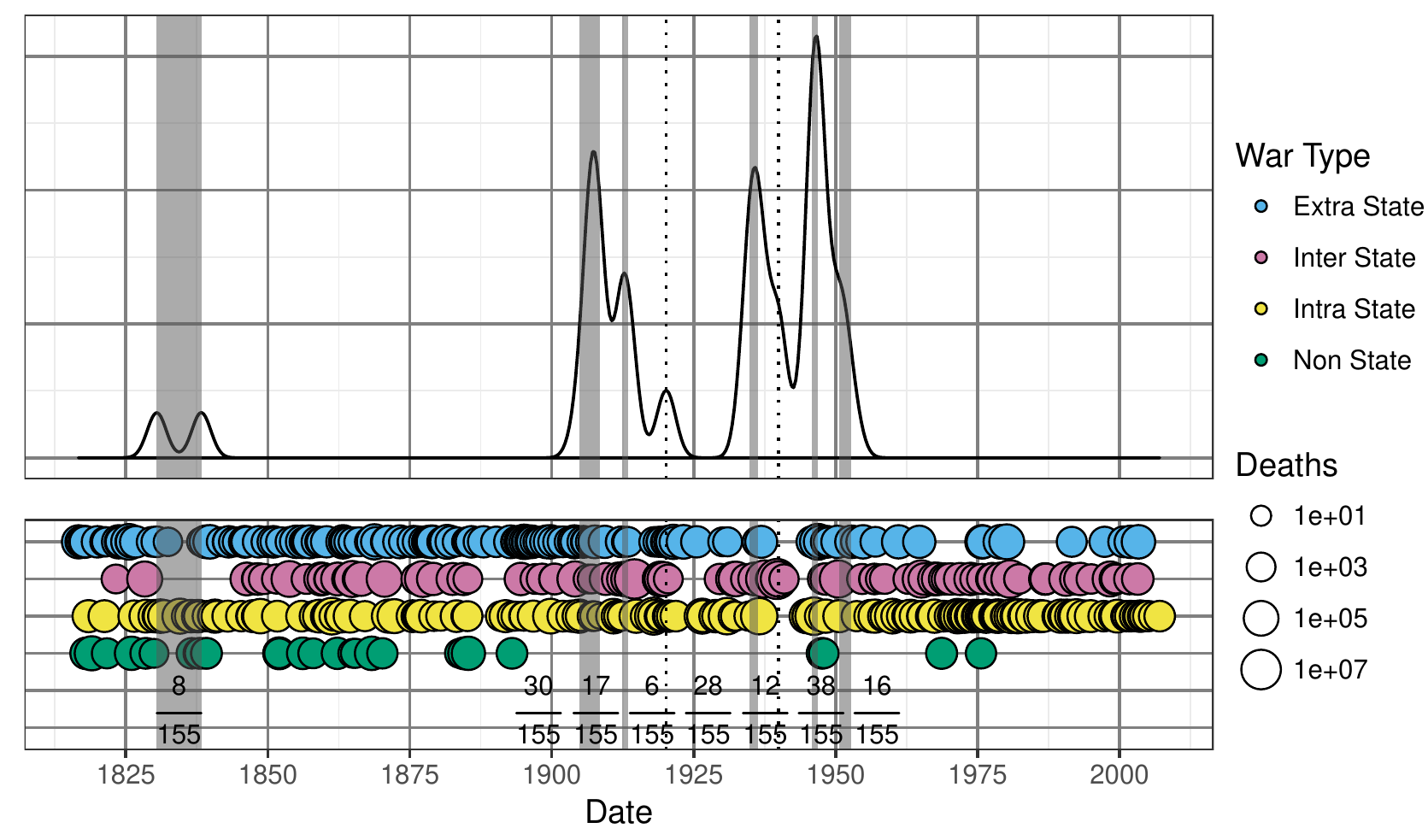}
       }\\
       \subfloat[CoW normalised data]{
           \includegraphics[width=0.85\linewidth]{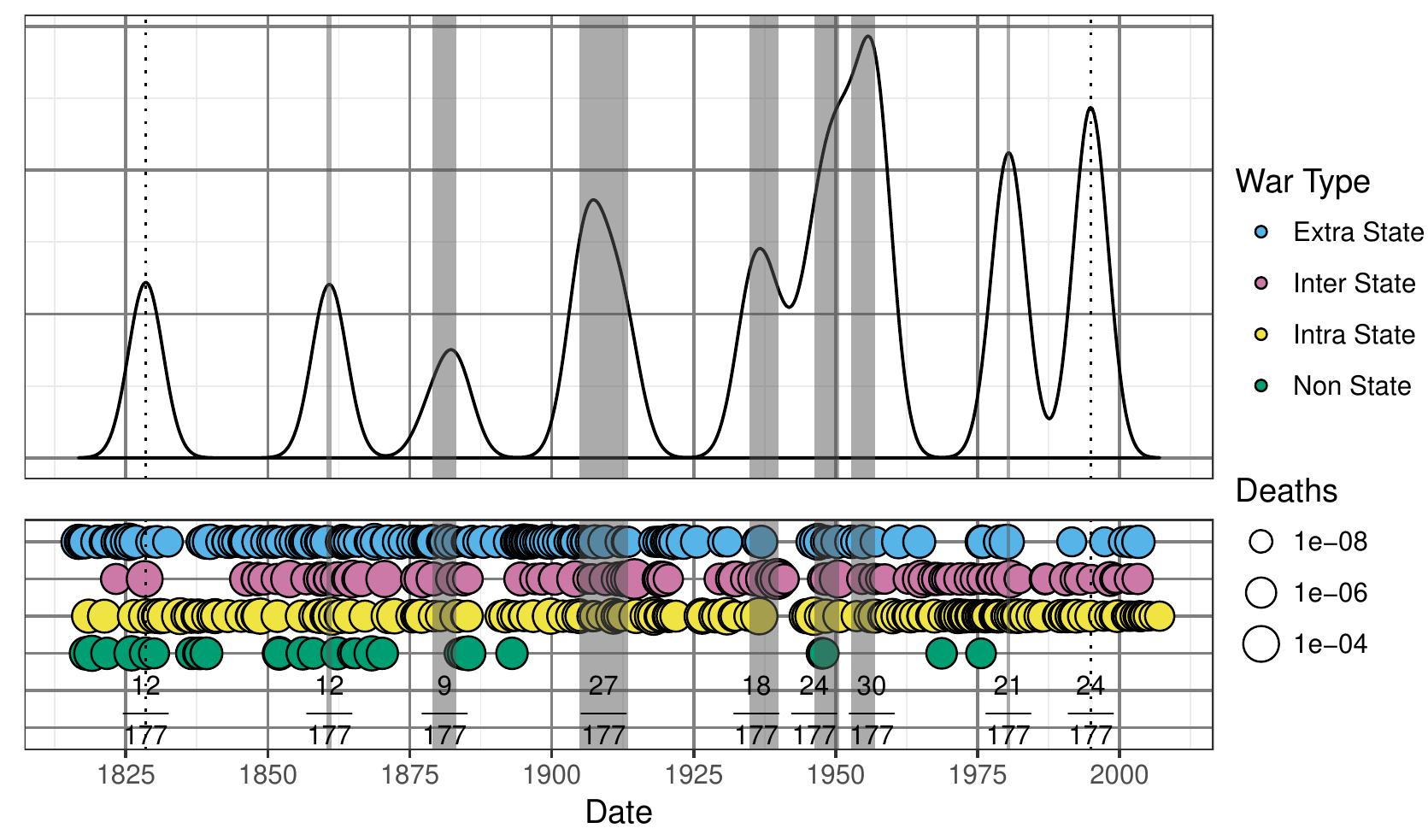}
       }
\end{figure}
\setcounter{figure}{11}
\begin{figure}
       \ContinuedFloat
       \centering
       \subfloat[Gleditsch raw data.]{
           \includegraphics[width=0.85\linewidth]{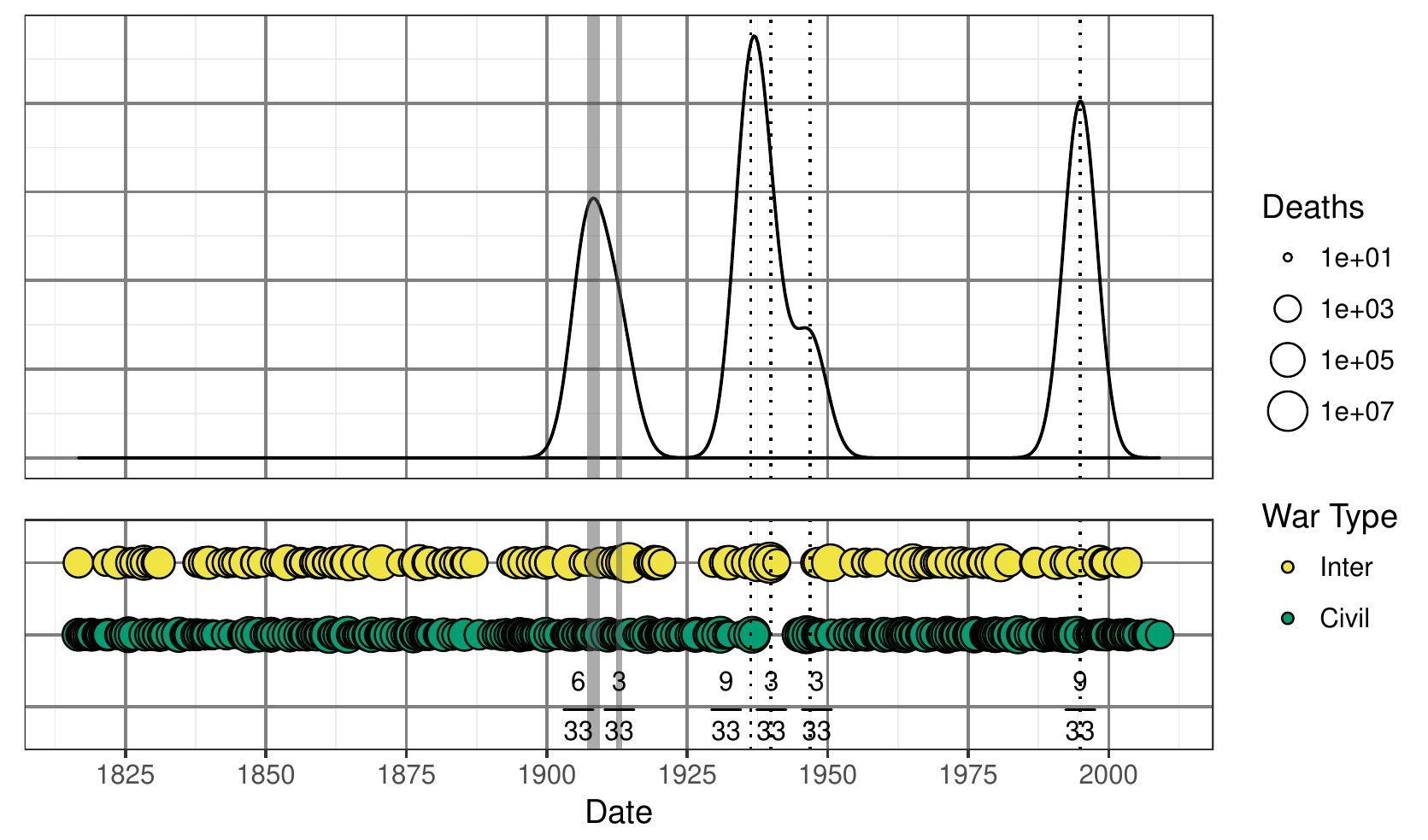}
       }\\
       \subfloat[Gleditsch normalised data.]{
           \includegraphics[width=0.85\linewidth]{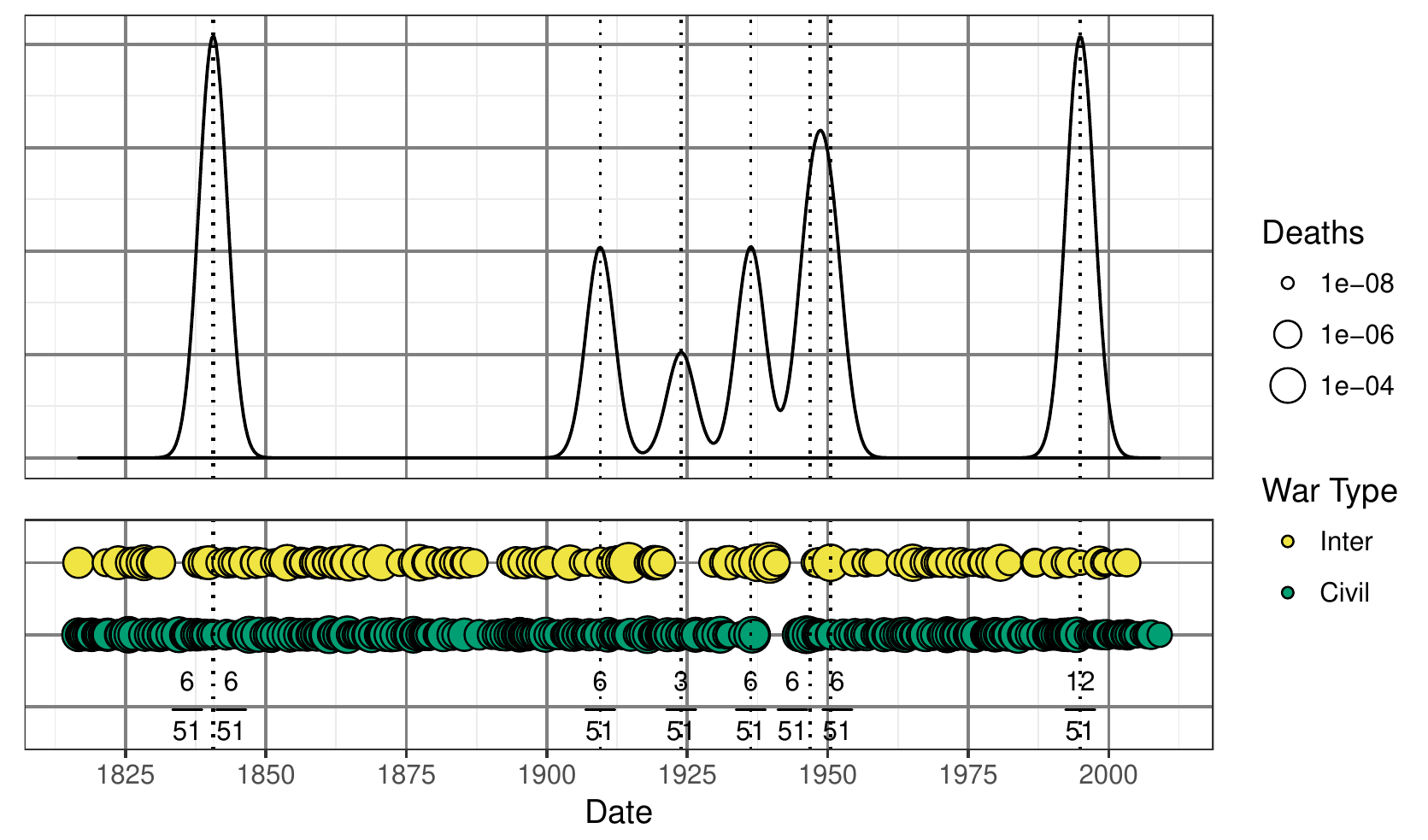}
       }
       \caption{\textbf{Results for internal meta-analyses performed on all changepoints found in any combination of subsets within the datasets.} In order, these plots correspond to (i) CoW raw data, (ii) CoW normalised data, (iii) Gleditsch raw data, and (iv) Gleditsch normalised data. In each plot, there are two images. The lower of each pair of images is a timeline of events occurring, sorted by subset. Above it is a density estimation of the locations of changepoints detected. The area under the curve of the estimate is proportional to the probability of finding a changepoint within that part of the dataset. Grey bars and dotted lines represent changepoints, in different locations or the same location respectively, that have been clustered. Numbers below the timelines indicate the fraction of identified changepoints so clustered.}
       \label{fig:meta}
\end{figure}

\section{Discussion}\label{sec:conclusion}
We have shown that recent advances in nonparametric changepoint analysis now allow for analysis of heavy-tailed data, an important class of data with unusual properties. Previous methods are prone to overfitting by comparison.
Our simulation study demonstrates that no single method fully captures the behaviour of heavy-tailed data, and we concluded that a combination of analyses more fully addressed the task of detecting changepoints.
In particular, we showed evidence for obtaining best segmentation results when combining ED-PELT \citep{Haynes17_Nonparametric} with CROPS \citep{Haynes17_CROPS} and mBIC \citep{Zhang07_Modified} penalties; moreover, this approach has the notable advantage of carrying no model-specific assumptions.

We emphasise that our approach is purely data-driven and we are explicitly not attempting to prove or disprove a particular thesis with our work.
The lively and fascinating debate about historical battle casualties has been hampered by disagreement over the existence and position of changepoints and the entanglement of the two strands of argument. In particular, the tendency within the literature to require that any putative changepoint be supported by an argument for its cause, and even in some cases to go looking for changepoints to support a hypothesis, creates a real danger of bias.
This leads to a number of issues, not least the potential for skewing the literature towards studies that find no changepoints.
In this context, it is nonetheless appropriate for us to speculate on possible reasons for the changepoints we have detected.

Applying our findings to historical battle deaths data, long considered power-law distributed \citep{Richardson60_Statistics,Clauset09_Power,Clauset18_Trends_and_Fluctuations}, revealed both new and old insights into how the data may have changed in time.
We detected the approximate beginning and end of the `great violence' 1910-1950 as changepoints, consistent with the idea that the World Wars marked a particularly violent period in human history.
We also observed possible changepoints in the 1800s and the 1990s across datasets and data presentations.
The former might indicate the change away from the so-called congress era, and the beginnings of the events that led to the revolutions in 1848.
The latter changepoint, around the end of the Cold War, supports the hypothesis put forward by \citet[see also \citet{Cederman17_Predicting}]{Gurr00_Ethnic}.

Our study provides a demonstration of a practical methodology, leveraging recent techniques to provide the best possible answer to whether changepoints exist in battle deaths data.
Additional rigour would require the development of changepoint detection techniques specifically designed for power-law distributions while retaining the ability to detect multiple changepoints.
Such distributions are of significant potential interest, including diverse areas such as blackouts, book sales, and terrorism \citep{Clauset09_Power}.
Furthermore we have not considered the possibility of continuous changes in underlying distributions over time such as those postulated by \citet{Pinker11_Better, Pinker18_Enlightenment}.
Our analysis takes an important step forward in answering whether changes exist, but stops short of integrating analysis of both continuous and discrete changes.
Nonetheless our study provides an essential statistical benchmark: driven by only the features of the data, we have demonstrated that the latest techniques show the existence of changepoints in well documented and publicly available datasets of battle deaths.

\clearpage
\bibliographystyle{rss}
\bibliography{cpa_combined_bib}
\end{document}